\documentclass[a4paper]{article}

\usepackage{INTERSPEECH2020}
\usepackage{multirow}
\usepackage{subfigure}
\usepackage{footnote}
\usepackage{hyperref}

\title{An End-to-end Architecture of Online Multi-channel Speech Separation }
\name{Jian Wu$^{1,2*}$, Zhuo Chen$^3$,Jinyu Li$^3$,Takuya Yoshioka$^3$, Zhili Tan$^2$, Ed Lin$^2$, Yi Luo$^3$, Lei Xie$^{1\dag}$}
\address{
  $^1$Audio, Speech and Language Processing Group (ASLP), School of Computer Science, \\ Northwestern Polytechnical University, Xi'an, China \\ \thanks{$^*$ Work done during internship at Microsoft STCA Beijing.} \thanks{$\dag$ Lei Xie is the corresponding author.}
  $^2$Microsoft, STCA, Beijing, China, 
  $^3$Microsoft, One Microsoft Way, Redmond, WA, USA }
\email{\{jianwu,lxie\}@nwpu-aslp.org, \{zhuc,jinyli,tayoshio\}@microsoft.com}

\begin{document}

\maketitle

\begin{abstract}

Multi-speaker speech recognition has been one of the key challenges in conversation transcription as it breaks the single active speaker assumption employed by most state-of-the-art speech recognition systems. Speech separation is considered as a remedy to this problem. Previously, we introduced a system, called \textit{unmixing}, \textit{fixed-beamformer} and \textit{extraction} (UFE), that was shown to be effective in addressing the speech overlap problem in conversation transcription. With UFE, an input mixed signal is processed by fixed beamformers, followed by a neural network post filtering. Although promising results were obtained, the system contains multiple individually developed modules, leading potentially sub-optimum performance.
In this work, we introduce an end-to-end modeling version of UFE. To enable gradient propagation all the way, an attentional selection module is proposed, where an attentional weight is learnt for each beamformer and spatial feature sampled over space.
Experimental results show that the proposed system achieves comparable performance in an offline evaluation with the original separate processing-based pipeline, while producing remarkable improvements in an online evaluation.

\end{abstract}

\section{Introduction}
\label{intro}
Deep learning approaches have brought about remarkable progress to speaker-independent speech separation in the past few years \cite{kolbaek2017multitalker,hershey2016deep,wang2019deep, luo2019conv}.
The separated signal quality has been steadily improved on benchmark 
datasets such as WSJ0-2mix~\cite{hershey2016deep}. However, multi-talker speech recognition still remains to be a challenging problem. 

Speech separation is a common practice to handle the speech overlaps. Existing efforts in overlapped 
speech recognition can be roughly categorized into two families: building a robust separation 
system as a front-end processor to automatic speech recognition (ASR) tasks
 \cite{higuchi2017online, zmolikova2017speaker,drude2017tight,chen2018multi,boeddeker2018front,wu2019improved,yoshioka2018recognizing} 
 or developing multi-talker aware ASR models 
 \cite{yu2017recognizing, chen2018progressive,lam2019extract,kanda2019auxiliary,chang2019end,delcroix2019end,k2020serialized}. Although 
 better performance can be expected from the end-to-end training including ASR, the 
independent front end processing approach is often preferable in real world applications 
 such as meeting transcription~\cite{PrincetonASRU2019} for two reasons. 
 Firstly, in the conversation transcription systems, the front end module benefits 
 multiple acoustic processing components, including speech recognition, diarization, and speaker verification.
 Secondly, commercial ASR models are usually trained with a tremendous amount of data and is highly engineered, making it extremely costly to change the training scheme.

The recent work of \cite{PrincetonASRU2019} applied speech separation to
a real-world conversation transcription task, where a multi-channel separation network, 
namely speech unmixing network, trained with permutation invariant training (PIT)~\cite{kolbaek2017multitalker} continuously separates 
the input audio stream into two channels, ensuring each output channel only contains at most one
activate speaker. A mask-based adaptive Minimum Variance Distortionless Response (MVDR) beamformer was used for generating enhanced signals. In \cite{yoshioka2019low}, a fixed beamformer based separation solution was introduced, namely the \textit{unmixing}, \textit{fixed-beamformer} and \textit{extraction} (UFE) system.
The mask-based adaptive beamformer of the speech unmixing 
system is replaced by a process selecting two fixed beamformers from a pre-defined set of beamformers by using a sound source localization (SSL) based beam selection algorithm. This is followed by the speech extraction model introduced in \cite{chen2018multi} to filter the residual interference in the selected beams.
The UFE system has comparable performance with MVDR-based approach, with reduced processing latency. 

One limitation of the UFE system lies in its modularized optimization, where each component is individually trained with an indirect objective function. For example, the signal reconstruction objective function used for speech unmixing does not necessarily benefit the accuracy of UFE's beam selection module.
As a subsequent work of \cite{yoshioka2019low}, in this paper, we propose a novel end-to-end structure of UFE (E2E-UFE) model, which utilizes a similar system architecture to UFE, with improved performance thanks to end-to-end optimization. To enable joint training, several updates are implemented on the speech unmixing and extraction networks. We also introduce an attentional module to allow the gradients to propagate though the beam selection module, which  was non-differentiable in the original UFE. The performance of the E2E-UFE is evaluated in both block online and offline setups.
Our experiments conducted on simulated and semi-real two-speaker mixtures show that E2E-UFE yields comparable results with the original UFE system in the offline evaluation. Significant WER reduction is observed in the block online evaluation.


\section{Overview of UFE System}
\label{sect2}

\begin{figure}[!tbp]
\centering
\includegraphics[width=0.45 \textwidth]{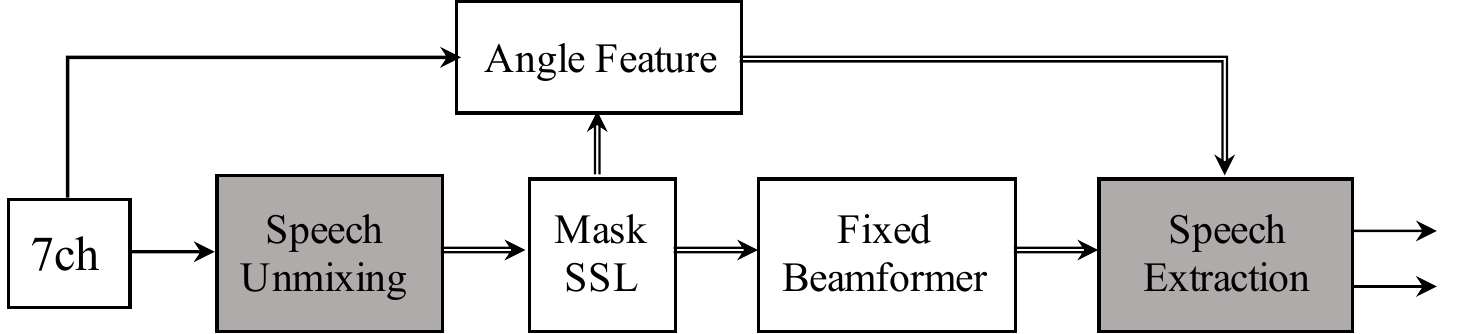}
\caption{Overview of the UFE system. The grey block is an neural network trained independently.}

\label{upe}
\end{figure}

The outline of the UFE pipeline is depicted in Figure. \ref{upe}, which consists 
four major components, the fixed beamformer, mask based sound source localization (SSL), 
speech unmixing network and location based speech extraction network. 

In UFE, the $M$-channel short-time Fourier transform (STFT) of the input speech mixture 
$\mathbf{Y}_{0, \cdots, M - 1} = \{\mathbf{Y}_{0}, \cdots, \mathbf{Y}_{M - 1}\}$ is firstly processed by the speech unmixing module, where a time-frequency mask (TF mask) is estimated for each 
participating speaker. In this work, we set the maximum number for simultaneously talking speakers to two so two masks $\mathbf{M}_{0,1} \in \mathbb{R}^{T \times F}$ are generated by unmixing network. 
The speech unmixing module is trained with permutation invariant training (PIT) criteria 
with scaled-invariant signal-to-noise ratio (Si-SNR)~\cite{le2019sdr} objective function:
\begin{equation}
    \mathcal{L} =  -\max_{\phi \in \mathcal{P}} \sum_{(i, j) \in \phi} \text{Si-SNR}(\mathbf{s}_i, \mathbf{x}_j),
    \label{pit}
\end{equation}
where $\mathcal{P}$ refers all possible permutations, $\mathbf{x}_j$ is the clean reference of speaker $j$, and $\mathbf{s}_i$ refers the separated signal of speaker $i$, which is obtained via inverse short-time Fourier transform (iSTFT):
\begin{equation}
    \mathbf{s}_i = \text{iSTFT} \left(\mathbf{M}_{i} \odot \mathbf{Y}_{0} \right).
\end{equation}


Then the sound source localization module is applied to estimate the spatial 
angle for each separated source with weighted maximum likelihood estimation \cite{yoshioka2019low}. The direction of the $i$-th speaker 
is estimated via finding a discrete angle $\theta$ sampled from 0$^\circ$ to 360$^\circ$ 
that maximizes the following function:
\begin{equation}
    \mathcal{D}_{\theta, i} = - \sum_{t,f} \mathbf{M}_{i,tf} \log \left(1 - \frac{\vert \mathbf{y}^H_{t,f} \mathbf{h}_{\theta, f} \vert^2}{1 + \epsilon} \right)
    \label{ssl}
\end{equation}
where $\mathbf{h}_{\theta, f}$ is the normalized steer vector on each frequency band $f$ for source 
direction $\theta$, $\epsilon$ refers a small flooring value, and $t$ denotes frame index in STFT.

With the estimated direction, one beamformer is then selected for each source from a set of pre-defined beamformer, defined as  $\mathbf{w}_{n,f} \in \mathbb{C}^{M \times 1}$,
where $n$ indexes the beam and each beam has an center angle that is sampled uniformly across the space, and the beamformed signal on each time-frequency bin is obtained by Eqn. \ref{bf}
\begin{equation}
   b_{i, t, f} = \mathbf{w}_{i,f}^H \mathbf{y}_{t,f},
   \label{bf}
\end{equation}
where $\mathbf{y}_{t,f} = [\mathbf{Y}_{0, tf}, \cdots, \mathbf{Y}_{M - 1, tf}]^T$.

Finally, the location based speech extraction~\cite{chen2018multi} is applied 
on each selected beam, and estimates the TF mask based on the input of the beam spectrogram, the inter-microphone phase difference (IPD) and the angle feature~\cite{chen2018multi,wang2018spatial}. The angle feature on frequency band $f$ is computed as
\begin{equation}
    \label{ang}
    \mathbf{a}_{\theta, f} = \frac{1}{P} \sum_{i,j \in \psi} \cos (\mathbf{o}_{ij,f} - \Delta_{\theta, ij, f}),
\end{equation}
where $\psi$ contains $P$ microphone pairs and and $\mathbf{o}_{ij,f} = \angle \mathbf{y}_{i,f} - \angle \mathbf{y}_{j,f}$ represents the observed IPD between channel $i$ and $j$. $\Delta_{\theta, ij, f}$ is the ground truth phase difference given the direction of arrival $\theta$ and array geometry. The final output signal is obtained via applying the TF masking on corresponding selected beam, followed by iSTFT.





As fixed beamformer doesn't need to estimate filter coefficients based on input data, it has the potential achieve low latency processing and more robust performance in challenge acoustic environments. And the speech extraction network compensates the limitation in spacial discrimination in fixed beamformer. 

\begin{figure*}[htb]
\centering
\includegraphics[width=0.85 \textwidth]{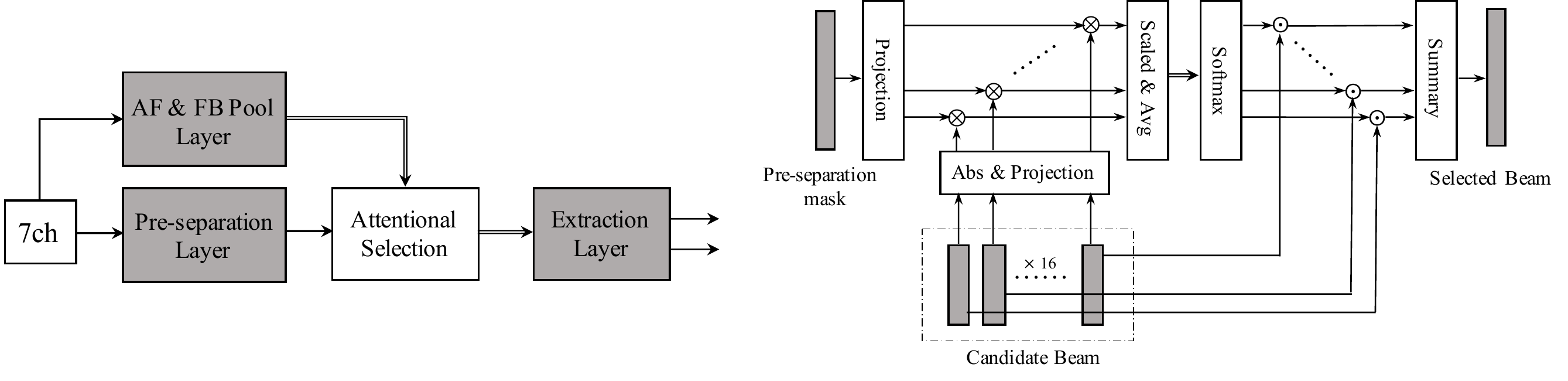}
\caption{Overview of the E2E-UFE system (left) and scheme of the attentional beam selection (right). 
AF and FB are abbreviation of the angle feature and fixed beamformer, respectively. The extraction layers accept both of the weighted beam and angle feature.}
\label{e2e-ufe}
\end{figure*}

\section{End-to-end UFE}
\label{sect3}

The proposed end-to-end UFE system is depicted in Figure \ref{e2e-ufe}. The overall system workflow is similar to original UFE, while the E2E framework largely simplifies the whole process. The proposed system takes the multi-channel recording as input, and directly outputs two separated speech. A single objective function on the top of the network is used to optimize all parameters.

In original UFE, three components are non-differentiable, which are SSL module, beam selection module and angle feature extraction. To ensure the joint training, we introduce updates to each component. 

\subsection{Pre-separation layer}

In E2E framework, the permutation ambiguity is handled in the final objective function, the unmixing module in UFE reduces to a stack of pre-separation layers. Same as the original UFE, the  network takes the IPD and spectrogram of first channel recording as input feature. The pre-separation layers consists of a stack of recurrent layers, followed by $H$ linear projection layers. Here we use $H = 2$ as we consider at most 2 speakers in this paper. After processed by pre-separation layers, an intermediate representation $\mathbf{E} \in \mathbb{R}^{H\times T \times K}$ is formed where $K$ denotes the embedding dimension. We refer $\mathbf{E}$ as the ``pre-separation mask'' in later context. 

\subsection{Attentional selection}

To avoid the hard angle selection in sound source module, i.e.  Eqn. \ref{ssl}. An attention module is applied in E2E-UFE system, which consists of a pool of beamformed signal and angle feature, followed by an attentional selection to estimate the location based bias for final extraction layers.

\subsubsection{Spatial feature pool}



The spatial feature pool is formed by stacking the spatial feature pointing to different directions. Two pools are formed, one for fixed beamforming and the other for angle feature. For beam pool $\mathbf{B} \in \mathbb{C} ^ {N_b\times T \times F}$, we calculate the spectrogram of signal obtained though all pre-defined fixed beamformer. In this work, we use $N_b = 18$ beamformers to scan the horizontal space, i.e., 20 degree is covered by each beamformer.  The angle feature pool $\mathbf{A} \in \mathbb{R}^{N_a \times T \times F}$ are formed similarly, with $N_a = 36$ directions. 

Note that in original UFE, only the beam and angle 
feature corresponding to the selected angle are calculated, while the E2E 
UFE calculates beamformed signal and angle feature from all directions beforehand, 
resulting in an increased computation burden. 
But this also open the possibility 
of jointly optimize the beamformer and the angle feature, as 
suggested in \cite{minhua2019frequency}, as they are now part of the network. 
In this work, we freeze the beamformer filter coefficients and angle 
feature representation. The complex operation in 
beamforming is implemented using the multiplication of two real 
matrices \cite{trabelsi2017deep}.





\subsubsection{Attentional beam \& angle selection}

With pre-separation mask $\mathbf{E}$, beam pool $\mathbf{B}$ 
and angle feature set $\mathbf{A}$ as input, an attention selection module 
is implemented to form the location based acoustic bias for each source. 
The intuition for the attention selection is straightforward, 
where one attention weight is estimated for each beam and angle feature, 
based on their learnt similarity with pre-separation representations, 
followed by a weighted sum to form the final beam and angle feature 
that are sent to the final extraction layers. In more detail, 
the attention module is operated in four steps. We use the beam attention 
as example for illustration, while the selection of angle feature operates 
in the same manner. The corresponding scheme is depicted in Figure. 2.

Firstly, the pre-separation mask and beam pool are projected using two linear 
layers:
\begin{align}
    \mathbf{V}^P & = \mathbf{E} \mathbf{W}_p, \\
    \mathbf{V}^B & = |\mathbf{B}| \mathbf{W}_b,
    \label{proj}
\end{align}
where $\mathbf{W}_p \in \mathbb{R}^{K\times D}$ and $\mathbf{W}_b\in \mathbb{R}^{F\times D}$ are projection layer weights that convert the pre-separation mask
and beam pool into the same dimension $D$, resulting updated embedding matrices
$\mathbf{V}^P \in \mathbb{R}^{H\times T \times D}$ and  
$\mathbf{V}^B \in \mathbb{R}^{N_b \times T \times D}$.

Then a pair-wised similarity matrix is defined between each 
frame in $\mathbf{V}^P$ and $\mathbf{V}^B$ using dot product distance, 
scaled by $(\sqrt{D})^{-1}$. Averaging the similarity matrix along the time 
axis resulted in beam selection for different time resolution, which is passed 
by the softmax function to generate the final weight. In Eqn. \ref{att}, $s_{h,b,t}$ is the similarity score between $h$-th pre-separation mask and $b$-th beam at time $t$, 
$\hat{s}_{h,b}$ refers the time averaged weights and $w_{h,b}$ is the final attention 
weight for each beam. Finally, the weight average operation is performed in order to get the combined beam $\hat{\mathbf{B}}_h$ for $h$-th speaker, as shown in the Eqn. 11.
\begin{align}
\label{att}
    s_{h,b,t} &= (\sqrt{D})^{-1} \left(\mathbf{V}^P_{h,t} \right)^T \mathbf{V}^B_{b,t}\\
    \hat{s}_{h,b} & = \left( T \right)^{-1} \sum_t s_{h,b,t}, \\
    w_{h,b} &= \text{softmax}_b(\hat{s}_{h,b}), \\
    \hat{\mathbf{B}}_h & = \sum_b w_{h,b} \mathbf{B}_b.
\end{align}

The combined angle feature $\hat{\mathbf{A}}_h$ can be calculated with the same 
mechanism. The proposed attention module connects the special feature, 
pre-separation and the later extraction step, ensuring the gradient can 
be passed in an end-to-end optimization scheme. Note that, the averaging 
step in Eqn. 9 can be adjusted according to different application scenarios. For offline processing, averaging over entire utterance usually leads to more robust estimation, assuming the position of the speaker is not changed. While averaging only based on past information is more desirable for online processing. 
The same mechanism can be applied with the other information as well, e.g., speaker 
inventory~\cite{wang2019speech} or visual clues, etc.

\subsection{Joint speech extraction}

The combined beam and angle feature estimated via the attentional selection module 
are processed by the extraction layers. The extraction layers have
essentially the same structure as the original UFE, except that the PIT training criteria 
is required as the permutation ambiguity is not disentangled by unmixing module in E2E framework. We use the clean source from the ground truth 
beam selection as training target, so both beam selection and wave reconstruction will be optimized with one objective function.
Denoting $\mathbf{r}_i$ as the training target for the speaker-$i$, the objective function is given in a permutation-free form:
\begin{equation}
    \mathcal{L} = -\max_{\phi \in \mathcal{P}} \sum_{(i,j) \in \phi} \text{Si-SNR}(\mathbf{s}_i, \mathbf{r}_j),
\end{equation}
where the $\mathbf{s}_i$ is the network's estimation of the speaker-$i$.

\section{Experiments}
\label{sect4}

\subsection{Dataset}

The proposed system was trained with multi-channel artificially mixed speech. A total of 1000 hours of training speech data was generated. Source clean speech signals were taken from publicly available datasets, including LibriSpeech\footnote{\href{http://www.openslr.org/12}{http://www.openslr.org/12/}}~\cite{panayotov2015librispeech}, Common Voice\footnote{\href{https://voice.mozilla.org/en}{https://voice.mozilla.org/en}}, as well as Microsoft internal recordings. Seven-channel signals were simulated by convolving clean speech signals with artificial room impulse responses (RIRs) generated with the image method~\cite{allen1979image}. We used the same microphone array geometry as the one used in \cite{yoshioka2019low}.
The T60 reverberation times were uniformly sampled from [0.1, 0.5] s with a room size of [2,20] m in length and width and [2,5] m in height. The speaker and microphone locations were randomly determined in the simulated rooms. Simulated isotropic noise~\cite{habets2007generating} was added to each mixing utterance at an SNR sampled from [10, 20] dB.  We made sure each speech mixture contained one or two speakers, with the mixing SNR between [-5, 5] dB and an average overlapping ratio of 50\%. All the data had a sampling rate of 16 kHz.

Two test sets were created for model evaluation. The first test set was created by using the same generation pipeline as the one for the training data, denoted as the \textit{simu} test set, which amounts to 3000 utterances. The speakers were sampled from the \textit{test-clean} set in LibriSpeech. There was no shared speakers in the training and test sets. The second test set was generated by directly mixing our internal real recorded multi-channel single speaker signals. 2000 mixed utterances were created with the same mixing strategy as in the training set, except that no scaling was applied on the source signals. We refer to this set as the \textit{semi-real} test set. For each set, we created two overlapping conditions, whose overlap ratio ranged from 20---50\% or 50--100\%. We denote these two condition as OV35 and OV75, respectively.

\subsection{Baseline systems}


The original UFE system served as the baseline of the proposed E2E architecture. We observed that when trained with the PIT criterion, the extraction model of UFE yielded significantly better results. Therefore, we used PIT-trained extraction in our UFE baseline. For reference, we included the results obtained with the fixed beamforming system applied directly to speech mixtures (Mixed Beam) and those obtained by applying the same beamformers to the clean utterances (Clean Beam), where the beams were selected based on oracle direction of arrival information.

\begin{table}[!t]
\centering
\caption{WER (\%) performance in the offline evaluation.}
\vspace{-1em}
\begin{tabular}{ccccc}
\toprule
\multirow{2}{*}{Method} & \multicolumn{2}{c}{\textit{simu}} & \multicolumn{2}{c}{\textit{semi-real}}  \\
& OV35 & OV75 & OV35 & OV75 \\  \midrule
Mixed Beam & 67.40 & 52.40 & 70.92 & 57.63 \\
Clean Beam & 10.67 & 10.56 & 20.34 & 19.71 \\ 
\midrule
UFE & 16.44 & 18.55  & 35.60 & 37.54 \\ \midrule
E2E-UFE & \textbf{16.85} & \textbf{18.98} & \textbf{33.89} & \textbf{35.92} \\ 
\bottomrule

\vspace{-0.9cm}
\end{tabular}
\end{table}

\subsection{Training scheme}


In the proposed E2E-UFE framework, both extraction and unmixing layers consisted of three contextual LSTM layers~\cite{li2019improving}, each with 512 nodes and a dropout rate of 0.2. For better convergence, the unmxing and extraction networks were pre-trained individually before joint optimization. The same model architecture for unmixing and extraction was used for the UFE baseline. The log magnitude spectrum with an FFT size of 512 and a hop of 256 samples was used as spectral features for all networks. For the unmixing network, cosIPDs between three microphone pairs $(1,4)$, $(2,5)$, $(3,6)$ were extracted.  

We used Adam optimizer and train both the networks for a maximum of 80 epochs with a weight decay value of $1e^{-5}$. The early stopping strategy was used to avoid over-fitting. Initial learning rate was set to $1e^{-3}$ and halved if no validation improvement was observed for two consecutive epochs. For joint training in E2E-UFE, a smaller learning rate $1e^{-4} $ was used for fine tuning.


\subsection{Evaluation scheme}

All systems were evaluated in offline and block online setups. 
In the offline evaluation, the system was allowed to use the information from an entire utterance. That is, SSL and attentional selection, i.e., Eqn. \ref{ssl} and \ref{att}, respectively, were performed by using averages over the whole utterance. 
In the block online processing, a double buffering~\cite{yoshioka2019low} scheme was applied, where each system estimated the output block-wisely through time. Each evaluation block contained a two second window, with additional two or four second history information. The hop between two evaluation block was two seconds, resulting in an average latency of one second.  


The word error rate (WER) was used as a performance metric. The ASR pipeline we used for decoding included a tri-gram language model and an acoustic model consisting of six layers of 512-element layer trajectory LSTM~\cite{li2018layer}. The acoustic model was trained with maximum mutual information (MMI)~\cite{vesely2013sequence} on 30k hours of noise-corrupted data.





\subsection{Results}

The offline evaluation results are shown in Table 1. The simple fixed beamforming (Mixed Beam) yielded a high WER even though it used the oracle DoA. The result of the clean beam sets the upper bound to the UFE performance. The proposed E2E-UFE system achieved comparable performance as the original UFE for the simulated data set, while demonstrating a clear performance advantage in \textit{semi-real} the semi-real set, showing the efficacy of the end-to-end training scheme. Overall, E2E-UFE achieved 4.8\% and 4.3\% relative WER reduction over the UFE system on OV35 and OV75 of the \textit{semi-real}, respectively, reaching 33.89\% and 35.92\% WERs.

Table 2 shows the block online evaluation results. E2E-UFE shows robustness  for different look-back configurations (a 2s or 4s history context),
achieving slightly worse results than for the offline evaluation on
both datasets. On the \textit{simu} set, E2E-UFE showed no significant degradation compared with the offline performance. It achieved lower WERs than the original UFE. On
the \textit{semi-real} set, it brought about a 12.47\% average relative
WER reduction compared with the UFE system using a 2 s history context, while on the \textit{simu} set, the relative reduction increases to 29.71\% . By contrast, the original UFE resulted in a much larger performance degradation for the online evaluation, degrading from 16.44/18.55\% to 24.10/31.40\% on the \textit{simu} set and 35.60/37.54\% to 44.05/43.15\% on the \textit{semi-real} set. One hypothesis for the robustness of E2E-UFE is that, during training, the E2E-UFE model
already optimized for wrong beam selections, 
while for the original UFE, only the correct beams were selected as input. Another potential reason could be that the sparification trick in \cite{yoshioka2019low} was not applied in either UFE or E2E-UFE, which might result in more energy leakage for UFE system, while E2E-UFE system doesn't suffer from this problem as all modules are jointly optimized.

\begin{table}[!t]
\centering
\caption{WER (\%) performance in the online evaluation.}
\vspace{-1em}
\begin{tabular}{ccccc}
\toprule
\multirow{2}{*}{Method (history)} & \multicolumn{2}{c}{\textit{simu}} & \multicolumn{2}{c}{\textit{semi-real}}  \\
& OV35 & OV70 & OV35 & OV70 \\  \midrule
UFE (2s) &  24.10 & 31.40 & 44.05 & 45.13 \\
UFE (4s) & 23.66 & 28.85 &  43.49 & 44.06 \\ \midrule
E2E-UFE (2s) & 17.50 & 19.43 &  38.64 & 39.98 \\
E2E-UFE (4s) & \textbf{17.09} & \textbf{19.10} & \textbf{36.67} & \textbf{39.11}\\ 

\bottomrule
\vspace{-0.9cm}
\end{tabular}
\end{table}
\vspace{-0.2cm}
\section{Conclusion}
In this paper, we proposed an end-to-end structure of multi-channel speech separation, named E2E-UFE, for robust ASR. It replaces the SSL module in the previously proposed UFE system with a small attention network and enables joint optimization of the unmixing and extraction networks. The experiments were conducted on two 2-speaker datasets (simulated and semi-real mixtures) and the performance was evaluated for both offline and online settings. The experimental results showed that E2E-UFE provided comparable performance with the UFE system in the offline situations and yielded an average relative WER reduction of 12.47\% 
on block online processing. 

\bibliographystyle{IEEEtran}

\bibliography{mybib}

\begin{thebibliography}{10}
\providecommand{\url}[1]{#1}
\csname url@samestyle\endcsname
\providecommand{\newblock}{\relax}
\providecommand{\bibinfo}[2]{#2}
\providecommand{\BIBentrySTDinterwordspacing}{\spaceskip=0pt\relax}
\providecommand{\BIBentryALTinterwordstretchfactor}{4}
\providecommand{\BIBentryALTinterwordspacing}{\spaceskip=\fontdimen2\font plus
\BIBentryALTinterwordstretchfactor\fontdimen3\font minus
  \fontdimen4\font\relax}
\providecommand{\BIBforeignlanguage}[2]{{%
\expandafter\ifx\csname l@#1\endcsname\relax
\typeout{** WARNING: IEEEtran.bst: No hyphenation pattern has been}%
\typeout{** loaded for the language `#1'. Using the pattern for}%
\typeout{** the default language instead.}%
\else
\language=\csname l@#1\endcsname
\fi
#2}}
\providecommand{\BIBdecl}{\relax}
\BIBdecl

\bibitem{kolbaek2017multitalker}
M.~Kolb{\ae}k, D.~Yu, Z.-H. Tan, and J.~Jensen, ``Multitalker speech separation
  with utterance-level permutation invariant training of deep recurrent neural
  networks,'' \emph{IEEE/ACM Transactions on Audio, Speech and Language
  Processing (TASLP)}, vol.~25, no.~10, pp. 1901--1913, 2017.

\bibitem{hershey2016deep}
J.~R. Hershey, Z.~Chen, J.~Le~Roux, and S.~Watanabe, ``Deep clustering:
  Discriminative embeddings for segmentation and separation,'' in \emph{2016
  IEEE International Conference on Acoustics, Speech and Signal Processing
  (ICASSP)}.\hskip 1em plus 0.5em minus 0.4em\relax IEEE, 2016, pp. 31--35.

\bibitem{wang2019deep}
Z.-Q. Wang, K.~Tan, and D.~Wang, ``Deep learning based phase reconstruction for
  speaker separation: A trigonometric perspective,'' in \emph{ICASSP 2019-2019
  IEEE International Conference on Acoustics, Speech and Signal Processing
  (ICASSP)}.\hskip 1em plus 0.5em minus 0.4em\relax IEEE, 2019, pp. 71--75.

\bibitem{luo2019conv}
Y.~Luo and N.~Mesgarani, ``Conv-tasnet: Surpassing ideal time--frequency
  magnitude masking for speech separation,'' \emph{IEEE/ACM Transactions on
  Audio, Speech, and Language Processing}, vol.~27, no.~8, pp. 1256--1266,
  2019.

\bibitem{higuchi2017online}
T.~Higuchi, N.~Ito, S.~Araki, T.~Yoshioka, M.~Delcroix, and T.~Nakatani,
  ``Online mvdr beamformer based on complex gaussian mixture model with spatial
  prior for noise robust asr,'' \emph{IEEE Transactions on Audio, Speech, and
  Language Processing}, vol.~25, no.~4, pp. 780--793, 2017.

\bibitem{zmolikova2017speaker}
K.~Zmolikova, M.~Delcroix, K.~Kinoshita, T.~Higuchi, A.~Ogawa, and T.~Nakatani,
  ``Speaker-aware neural network based beamformer for speaker extraction in
  speech mixtures.'' in \emph{Interspeech}, 2017, pp. 2655--2659.

\bibitem{drude2017tight}
L.~Drude and R.~Haeb-Umbach, ``Tight integration of spatial and spectral
  features for bss with deep clustering embeddings.'' in \emph{Interspeech},
  2017, pp. 2650--2654.

\bibitem{chen2018multi}
Z.~Chen, X.~Xiao, T.~Yoshioka, H.~Erdogan, J.~Li, and Y.~Gong, ``Multi-channel
  overlapped speech recognition with location guided speech extraction
  network,'' in \emph{2018 IEEE Spoken Language Technology Workshop
  (SLT)}.\hskip 1em plus 0.5em minus 0.4em\relax IEEE, 2018, pp. 558--565.

\bibitem{boeddeker2018front}
C.~Boeddeker, J.~Heitkaemper, J.~Schmalenstroeer, L.~Drude, J.~Heymann, and
  R.~Haeb-Umbach, ``Front-end processing for the chime-5 dinner party
  scenario,'' in \emph{CHiME5 Workshop, Hyderabad, India}, 2018.

\bibitem{wu2019improved}
J.~Wu, Y.~Xu, S.-X. Zhang, L.-W. Chen, M.~Yu, L.~Xie, and D.~Yu, ``Improved
  speaker-dependent separation for chime-5 challenge,'' \emph{arXiv preprint
  arXiv:1904.03792}, 2019.

\bibitem{yoshioka2018recognizing}
T.~Yoshioka, H.~Erdogan, Z.~Chen, X.~Xiao, and F.~Alleva, ``Recognizing
  overlapped speech in meetings: A multichannel separation approach using
  neural networks,'' \emph{arXiv preprint arXiv:1810.03655}, 2018.

\bibitem{yu2017recognizing}
D.~Yu, X.~Chang, and Y.~Qian, ``Recognizing multi-talker speech with
  permutation invariant training,'' \emph{arXiv preprint arXiv:1704.01985},
  2017.

\bibitem{chen2018progressive}
Z.~Chen, J.~Droppo, J.~Li, and W.~Xiong, ``Progressive joint modeling in
  unsupervised single-channel overlapped speech recognition,'' \emph{IEEE
  Transactions on Audio, Speech, and Language Processing}, vol.~26, no.~1, pp.
  184--196, 2018.

\bibitem{lam2019extract}
M.~W. Lam, J.~Wang, X.~Liu, H.~Meng, D.~Su, and D.~Yu, ``Extract, adapt and
  recognize: an end-to-end neural network for corrupted monaural speech
  recognition,'' \emph{Proc. Interspeech 2019}, pp. 2778--2782, 2019.

\bibitem{kanda2019auxiliary}
N.~Kanda, S.~Horiguchi, R.~Takashima, Y.~Fujita, K.~Nagamatsu, and S.~Watanabe,
  ``Auxiliary interference speaker loss for target-speaker speech
  recognition,'' \emph{arXiv preprint arXiv:1906.10876}, 2019.

\bibitem{chang2019end}
X.~Chang, Y.~Qian, K.~Yu, and S.~Watanabe, ``End-to-end monaural multi-speaker
  asr system without pretraining,'' in \emph{ICASSP 2019-2019 IEEE
  International Conference on Acoustics, Speech and Signal Processing
  (ICASSP)}.\hskip 1em plus 0.5em minus 0.4em\relax IEEE, 2019, pp. 6256--6260.

\bibitem{delcroix2019end}
M.~Delcroix, S.~Watanabe, T.~Ochiai, K.~Kinoshita, S.~Karita, A.~Ogawa, and
  T.~Nakatani, ``End-to-end speakerbeam for single channel target speech
  recognition,'' \emph{Proc. Interspeech 2019}, pp. 451--455, 2019.

\bibitem{k2020serialized}
N.~Kanda, Y.~Gaur, X.~Wang, Z.~Meng, and T.~Yoshioka, ``Serialized output
  training for end-to-end overlapped speech recognition,'' \emph{arXiv preprint
  arXiv:2003.12687}, 2020.

\bibitem{PrincetonASRU2019}
T.~Yoshioka, I.~Abramovski \emph{et~al.}, ``Advances in online audio-visual
  meeting transcription,'' in \emph{Proc. Worksh. Automat. Speech Recognition,
  Understanding}, 2019.

\bibitem{yoshioka2019low}
T.~Yoshioka, Z.~Chen, C.~Liu, X.~Xiao, H.~Erdogan, and D.~Dimitriadis,
  ``Low-latency speaker-independent continuous speech separation,'' in
  \emph{ICASSP 2019-2019 IEEE International Conference on Acoustics, Speech and
  Signal Processing (ICASSP)}.\hskip 1em plus 0.5em minus 0.4em\relax IEEE,
  2019, pp. 6980--6984.

\bibitem{le2019sdr}
J.~Le~Roux, S.~Wisdom, H.~Erdogan, and J.~R. Hershey, ``Sdr--half-baked or well
  done?'' in \emph{ICASSP 2019-2019 IEEE International Conference on Acoustics,
  Speech and Signal Processing (ICASSP)}.\hskip 1em plus 0.5em minus
  0.4em\relax IEEE, 2019, pp. 626--630.

\bibitem{wang2018spatial}
Z.-Q. Wang and D.~Wang, ``On spatial features for supervised speech separation
  and its application to beamforming and robust asr,'' in \emph{2018 IEEE
  International Conference on Acoustics, Speech and Signal Processing
  (ICASSP)}.\hskip 1em plus 0.5em minus 0.4em\relax IEEE, 2018, pp. 5709--5713.

\bibitem{minhua2019frequency}
W.~Minhua, K.~Kumatani, S.~Sundaram, N.~Str{\"o}m, and B.~Hoffmeister,
  ``Frequency domain multi-channel acoustic modeling for distant speech
  recognition,'' in \emph{ICASSP 2019-2019 IEEE International Conference on
  Acoustics, Speech and Signal Processing (ICASSP)}.\hskip 1em plus 0.5em minus
  0.4em\relax IEEE, 2019, pp. 6640--6644.

\bibitem{trabelsi2017deep}
C.~Trabelsi, O.~Bilaniuk, Y.~Zhang, D.~Serdyuk, S.~Subramanian, J.~F. Santos,
  S.~Mehri, N.~Rostamzadeh, Y.~Bengio, and C.~J. Pal, ``Deep complex
  networks,'' \emph{arXiv preprint arXiv:1705.09792}, 2017.

\bibitem{wang2019speech}
P.~Wang, Z.~Chen, X.~Xiao, Z.~Meng, T.~Yoshioka, T.~Zhou, L.~Lu, and J.~Li,
  ``Speech separation using speaker inventory,'' in \emph{2019 IEEE Automatic
  Speech Recognition and Understanding Workshop (ASRU)}.\hskip 1em plus 0.5em
  minus 0.4em\relax IEEE, 2019, pp. 230--236.

\bibitem{panayotov2015librispeech}
V.~Panayotov, G.~Chen, D.~Povey, and S.~Khudanpur, ``Librispeech: an asr corpus
  based on public domain audio books,'' in \emph{2015 IEEE International
  Conference on Acoustics, Speech and Signal Processing (ICASSP)}.\hskip 1em
  plus 0.5em minus 0.4em\relax IEEE, 2015, pp. 5206--5210.

\bibitem{allen1979image}
J.~B. Allen and D.~A. Berkley, ``Image method for efficiently simulating
  small-room acoustics,'' \emph{The Journal of the Acoustical Society of
  America}, vol.~65, no.~4, pp. 943--950, 1979.

\bibitem{habets2007generating}
E.~A. Habets and S.~Gannot, ``Generating sensor signals in isotropic noise
  fields,'' \emph{The Journal of the Acoustical Society of America}, vol. 122,
  no.~6, pp. 3464--3470, 2007.

\bibitem{li2019improving}
J.~Li, L.~Lu, C.~Liu, and Y.~Gong, ``Improving layer trajectory lstm with
  future context frames,'' in \emph{ICASSP 2019-2019 IEEE International
  Conference on Acoustics, Speech and Signal Processing (ICASSP)}.\hskip 1em
  plus 0.5em minus 0.4em\relax IEEE, 2019, pp. 6550--6554.

\bibitem{li2018layer}
J.~Li, C.~Liu, and Y.~Gong, ``Layer trajectory lstm,'' \emph{arXiv preprint
  arXiv:1808.09522}, 2018.

\bibitem{vesely2013sequence}
K.~Vesel{\`y}, A.~Ghoshal, L.~Burget, and D.~Povey, ``Sequence-discriminative
  training of deep neural networks.'' in \emph{Interspeech}, vol. 2013, 2013,
  pp. 2345--2349.

\end{thebibliography}


\end{document}